%% file: arxiv.tex
\documentclass[final]{article}

\usepackage{arxiv}

\usepackage[utf8]{inputenc} 
\usepackage[T1]{fontenc}    
\usepackage{hyperref}       
\usepackage{url}            
\usepackage{booktabs}       
\usepackage{amsfonts}       
\usepackage{nicefrac}       
\usepackage{microtype}      
\usepackage{lipsum}
\usepackage{graphicx}

\usepackage{xcolor}
\usepackage{tikz}
\usepackage{amsmath}
\usepackage{amssymb}

\usepackage{lineno}
\usepackage{algorithm2e}
\usepackage{textcomp} 
\usepackage{subcaption}
\usepackage{adjustbox}

\usepackage{longtable}

\modulolinenumbers[5]
\usetikzlibrary{calc}
\usetikzlibrary{intersections}
\usetikzlibrary{through}

\title{Girasol, a Sky Imaging and Global Solar Irradiance Dataset}

\author{
  Guillermo Terr\'en-Serrano \\
  Department of Electrical and Computer Engineering \\
  The University of New Mexico \\
  Albuquerque, NM 87131, United States\\
  \texttt{guillermoterren@unm.edu} \\
  \And
  Adnan Bashir \\
  Department of Computer Science \\
  The University of New Mexico \\
  Albuquerque, NM 87131, United States\\
  \texttt{abashir@unm.edu} \\
   \And
  Trilce Estrada \\
  Department of Computer Science \\
  The University of New Mexico \\
  Albuquerque, NM 87131, United States\\
  \texttt{trilce@unm.edu} \\
 \And
  Manel Mart\'inez-Ram\'on \\
  Department of Electrical and Computer Engineering \\
  The University of New Mexico \\
  Albuquerque, NM 87131, United States\\
  \texttt{manel@unm.edu} \\
}

\begin{document}

\maketitle

\begin{abstract}
    The energy available in a microgrid that is powered by solar energy is tightly related to the weather conditions at the moment of generation. A very short-term forecast of solar irradiance provides the microgrid with the capability of automatically controlling the dispatch of energy. We propose a dataset to forecast Global Solar Irradiance (GSI)  using a data acquisition system (DAQ) that simultaneously records sky imaging and GSI measurements, with the objective of extracting features from clouds and use them to forecast the power produced by a Photovoltaic (PV) system. The DAQ system is nicknamed the \emph{Girasol Machine} (Girasol means Sunflower in Spanish). The sky imaging system consists of a longwave infrared (IR) camera and a visible (VI) light camera with a fisheye lens attached to it. The cameras are installed inside a weatherproof enclosure that it is mounted on a solar tracker. The tracker updates its pan and tilt every second using a solar position algorithm to maintain the Sun in the center of the IR and VI images. A pyranometer is situated on a horizontal mount next to the DAQ system to measure GSI. The dataset, composed of IR images, VI images, GSI measurements, and the Sun's positions, has been tagged with timestamps.
\end{abstract}


\section{Value of the Data}


\begin{itemize}
    \itemsep=0pt
    \parsep=0pt
    \item This dataset was generated using a sky imaging DAQ system for solar forecasting \cite{YANG2018}. This system is innovative in that it is mounted on a solar tracker so the Sun stays in the center of the images, and that it combines ground-based IR and VI light fisheye images \cite{MAMMOLI2019}. IR cameras have been previously used but not mounted on a solar tracker \cite{DIAGNE2013}.
    \item This dataset was originally conceived to address the problems related to very short-term forecasting of solar irradiance \cite{CHENG2017b}. In very short-term forecasting of solar irradiance the cloud information in the images is useful to accurately predict when the Sun may be occluded by a cloud. In these events, a shadow is projected over a PV system producing a drop in the energy supply \cite{FURLAN2012, KONG2020}. A GSI forecasting algorithm will provide the grid with the capability of managing the energy resources \cite{CHOU2019}.
    \item This dataset includes images captured using two different light sensors equipped with lenses. The information provided by the sensors may be combined to increase the diversity of features extracted from the clouds \cite{Deng2019}. Image processing is an important factor in the performance of a solar forecasting algorithm \cite{Ye2019}.
    \item In the context of machine learning and image processing, the feature extraction algorithm may be adapted to different applications in solar problems \cite{VOYANT2017, GARCIA2018}. For instance, the solar forecasting horizon can be changed depending on the application. The dataset can be used to forecast the effect of the clouds in thermosolar energy generation systems such as concentrated solar power \cite{CRESPI2018, CHEN2020}.
\end{itemize}

\section{Specifications table}
\input{revised_table}

\section{Data Description}

The repository contains recordings of the solar cycle from 242 days of 3 years. The total amount of data is 119GB. The sampling interval of the cameras is 15 seconds and the observation period is when the Sun's elevation angle is higher than $15^\circ$. There are approximately 1,200 to 2,400 captures per day from each camera depending on the day of the year.

\begin{itemize}
    \item VI images are 16 bits with a resolution of 450 $\times$ 450, intensity channel only. Approx. 240KB per frame. Between 200 MB to 400 MB per day depending on the amount of images in the directory. The images are saved in a lossless png format in the directory */visible. The images are named by the UNIX time in seconds.
    
    \item IR images are 16 bits with a resolution of 60 $\times$ 80. Approx. 8KB per frame. 20MB per day roughly constant. Images are saved in lossless png format in the directory */infrared. The image are named by the UNIX time in seconds.
    
    \item The pyranometer is sampled from 4 to 6 times per second. The measurements are saved in the directory */pyranometer in .csv files named with their date (\texttt{yyyy\_mm\_dd}), approx. 4,500KB to 7,500KB per file. The files contain UNIX time in the first column and GSI in $W/m^2$ in the second column. 
 
    \item The Sun position files are generated from the time recorded in the pyranometer files. The positions are saved in the directory */sun\_position in .csv files named with their date (\texttt{yyyy\_mm\_dd}), approx. 6,500KB to 11,500KB per file. The files contain the UNIX time in the first column, the elevation angle in the second column, and the azimuth angle in the third column, all of them computed with full precision UNIX time.
    
    \item The weather station sample interval is 10 minutes. Linear interpolation was applied to match the sampling interval of the pyranomter. The weather station files are saved in the directory */weather\_station in .csv files named with their date (\texttt{yyyy\_mm\_dd}), approx. 14.3MB top 25.4MB per file. The files are organized by columns which contain, from left to right: UNIX time, temperature in $^\circ C$, dew point in $^\circ C$, atmospheric pressure in $mmHg$, wind direction in $radians$, wind velocity in $mile/s$ and relative humidity in $\%$.
    
\end{itemize}

\section{Experimental Design, Materials and Methods}

The \emph{Girasol Machine} is composed of data acquisition hardware and software. In this section, we first summarize the specifications of the hardware. The described parts of the hardware are: the IR camera, the VI camera, the solar tracker and the pyranometer. We later describe the formulation of the algorithms in the software. These algorithms are designed to compute the Sun's position, to attenuate the noise in IR and VI images, and to fuse multiple exposure VI images together.

\subsection{Data Acquisition Hardware}

We built a system composed of VI and IR solar radiation cameras, a tracking system, and a pyranometer, to collect the data. A weatherproof enclosure contains the two USB cameras, and it is mounted on top of two servomotors. The cameras are connected to a motherboard running a solar tracking algorithm in parallel. This motherboard is placed inside of a different weatherproof enclosure together with a router, a servomotor control unit, a data storage system, and the respective power supplies. The enclosure’s degree of protection is IP66. This is an international standard utilized for electronic equipment that provides protection against dust and water (see Fig \ref{fig:interior}-\ref{fig:exterior}).

\subsubsection{Infrared Sensor}

\begin{figure}[!htb]
    \begin{subfigure}{0.3275\textwidth}
        \centering
        \includegraphics[scale = 0.34]{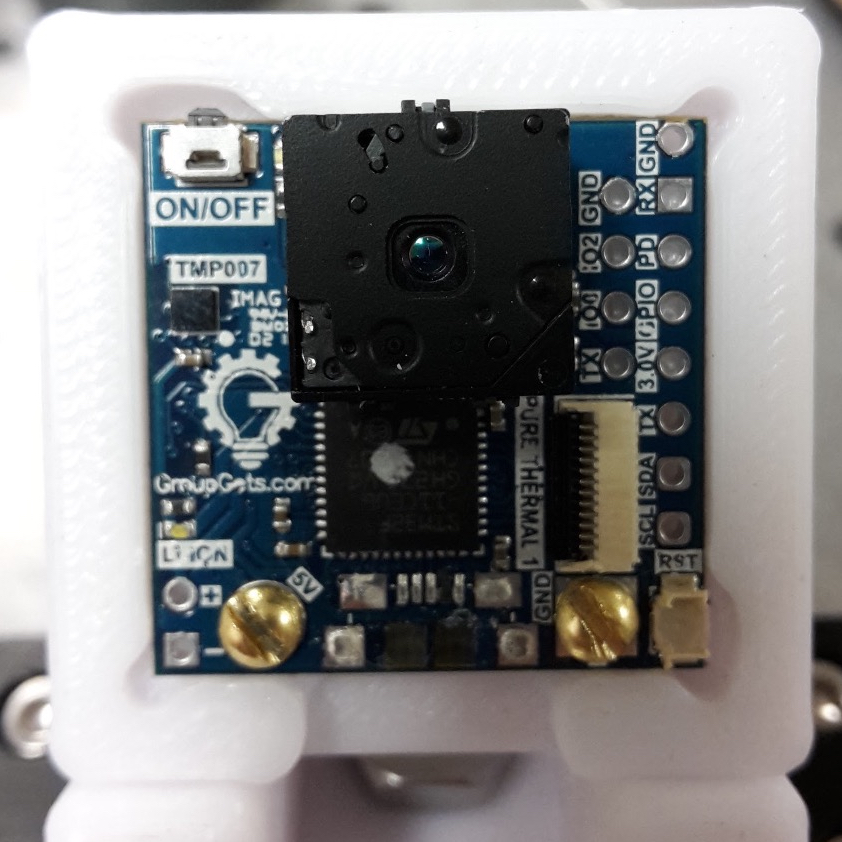}
    \end{subfigure}
    \begin{subfigure}{0.3275\textwidth}
        \centering
        \includegraphics[scale = 0.0875]{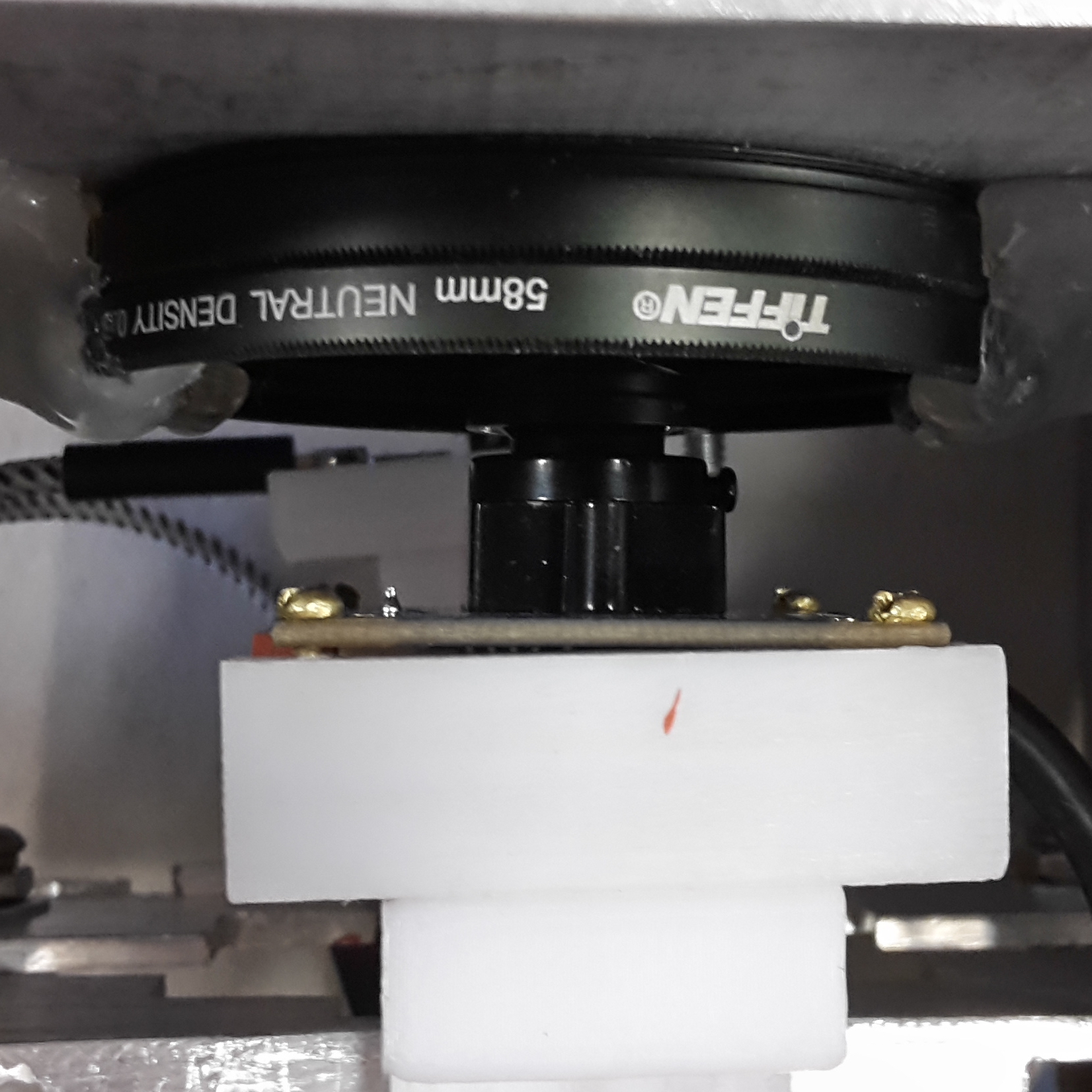}
    \end{subfigure}
    \begin{subfigure}{0.3275\textwidth}
        \centering
        \includegraphics[scale = 2.]{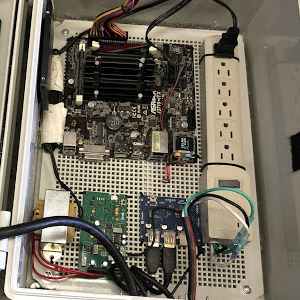}
    \end{subfigure}
    \caption{Details of IR (left) and VI (middle) cameras screwed onto their individual 3D printed support, inside the custom made enclosure, and placed in front of their respective apertures. The supports are fixed to the enclosure's structured by means of adjustable clamps. The junction box (right) contains the computer and controllers.}
    \label{fig:interior}
\end{figure}

The IR sensor used to capture the images is a FLIR$\copyright$ Lepton 2.5 camera\footnote{http://www.flir.com} which is mounted on a Pure Thermal 1 board\footnote{https://groupgets.com} manufactured and distributed by Group Gets. The Lepton 2.5 sensor produces thermal images by measuring long-wave infrared (see Fig. \ref{fig:interior}). It captures IR radiation with a nominal response wavelength band from 8 to 14 $\mu m$. 
	
The dimensions of a Lepton 2.5 are 8.5 $\times$ 11.7 $\times$ 5.6 mm. It has $51^\circ$ horizontal Field Of View (FOV) and $63.5^\circ$ diagonal FOV with type f1.1 silicon doublet lens. The resolution is 80 (horizontal) $\times$ 60 (vertical) active pixels. The thermal sensitivity is $<50$ mK. The camera integrates digital thermal image processing functions that include automatic thermal environment compensation, noise filtering, non-uniformity correction and gain control. This sensor can produce or stream an image in $<0.5$ seconds. It operates at 150 mW nominal power, and has a low power standby mode. The Pure Thermal 1 board adds functionalities to the IR sensor such as thermal video transmission over USB, which works with a USB Video Class (UVC) library on Windows, Linux, Mac and Android. It also includes on-board image processing without the need of an external system. The microprocessor is an STM32F411CEU6 ARM. It also includes a contactless thermopile  temperature sensor to manually calibrate a Lepton module. The operational temperatures for Flat Field Correction (FFC) function range from $-10^\circ$C to $65^\circ$C. The captured images are digitized in Y16 format. It is a single channel format that only quantifies intensity levels.

\subsubsection{Visible Sensor}

The VI camera is a 5 megapixel color sensor with USB 2.0 manufactured by ELP$\copyright$. The sensor is an OV5640 with maximum resolution of 2592 (horizontal) $\times$ 1944 (vertical) pixels and $170^\circ$ FOV with a fisheye type lens that is adjustable within the range of 2.1 to 6 mm (see Fig. \ref{fig:interior}). The pixel size is 1.4 $\times$ 1.4 $\mu m$, and the image area is 3673.6 $\times$ 2738.4 $\mu m $. The image stream rate is 30 frames per second at 640 $\times$ 480 pixel resolution. The communication protocol is UVC and its interface is a USB 2.0 high speed. The dynamic range is 68 dB, and it has a mount-in shutter that can control the frame exposure time. The camera-board has built-in functions that can be enabled for Automatic Gain Control (AGC), Automatic Exposure Control (AEC), Automatic White Balance (AWB) and Automatic Black Focus (ABF). The sensor's brightness, contrast, hue, saturation, sharpness, gamma, white balance, exposure and focus are software adjustable. The power consumption in VGA resolution is 150mW. The dimensions of the camera board are 38 x 38 mm. The recommended operational temperatures for stable images range from $0^\circ$C to $60^\circ$C. The output format of the camera is YUYV. YUYV is a 3 channel format in the YUV color space, where Y represents brightness, U blue color projection and V red color projection. UVC drivers can operate in Windows, Linux, MAC and Android.

\begin{figure}[!htb]
    \begin{subfigure}{0.3275\textwidth}
        \centering
        \includegraphics[scale = 0.08]{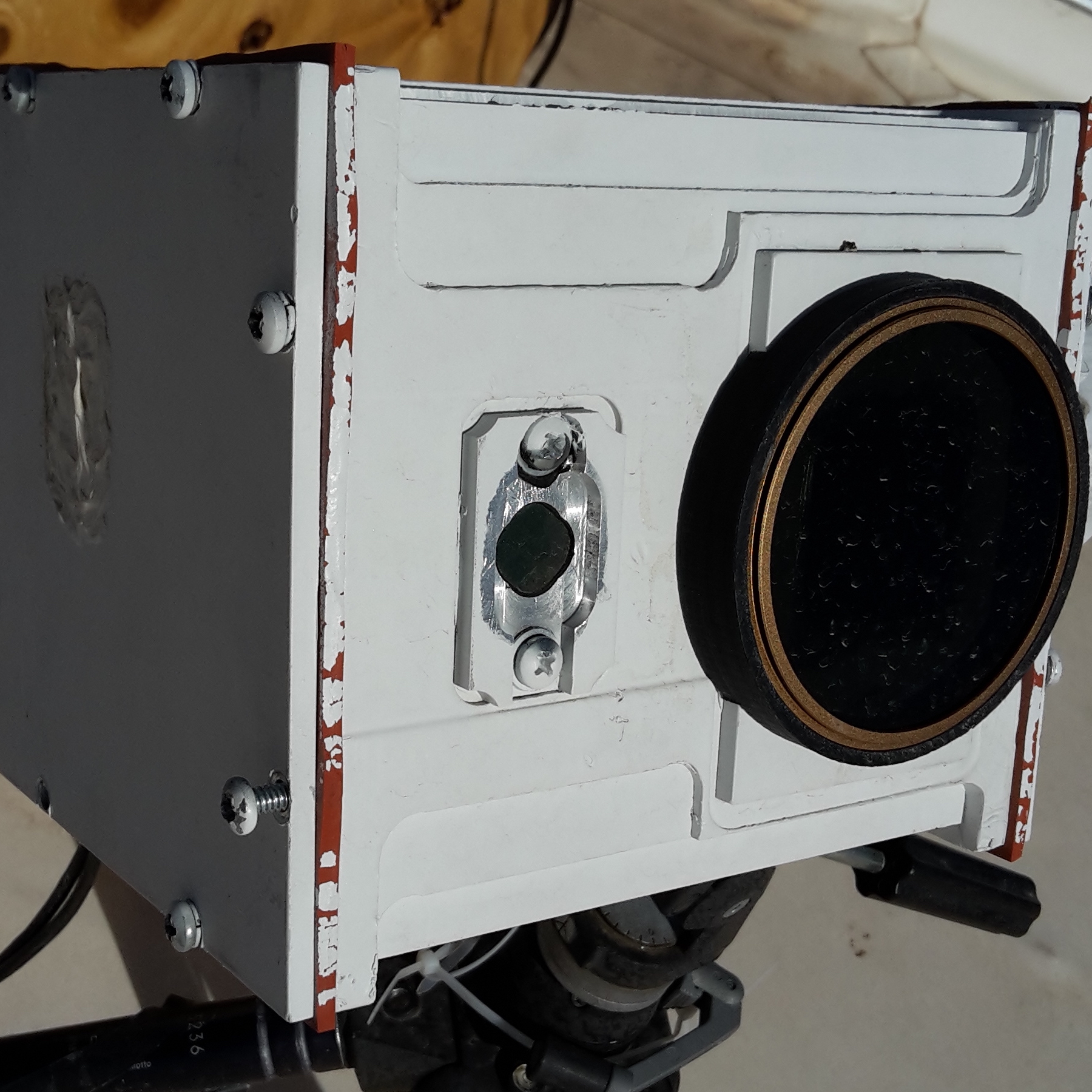}
    \end{subfigure}
    \begin{subfigure}{0.3275\textwidth}
        \centering
        \includegraphics[scale = 0.08]{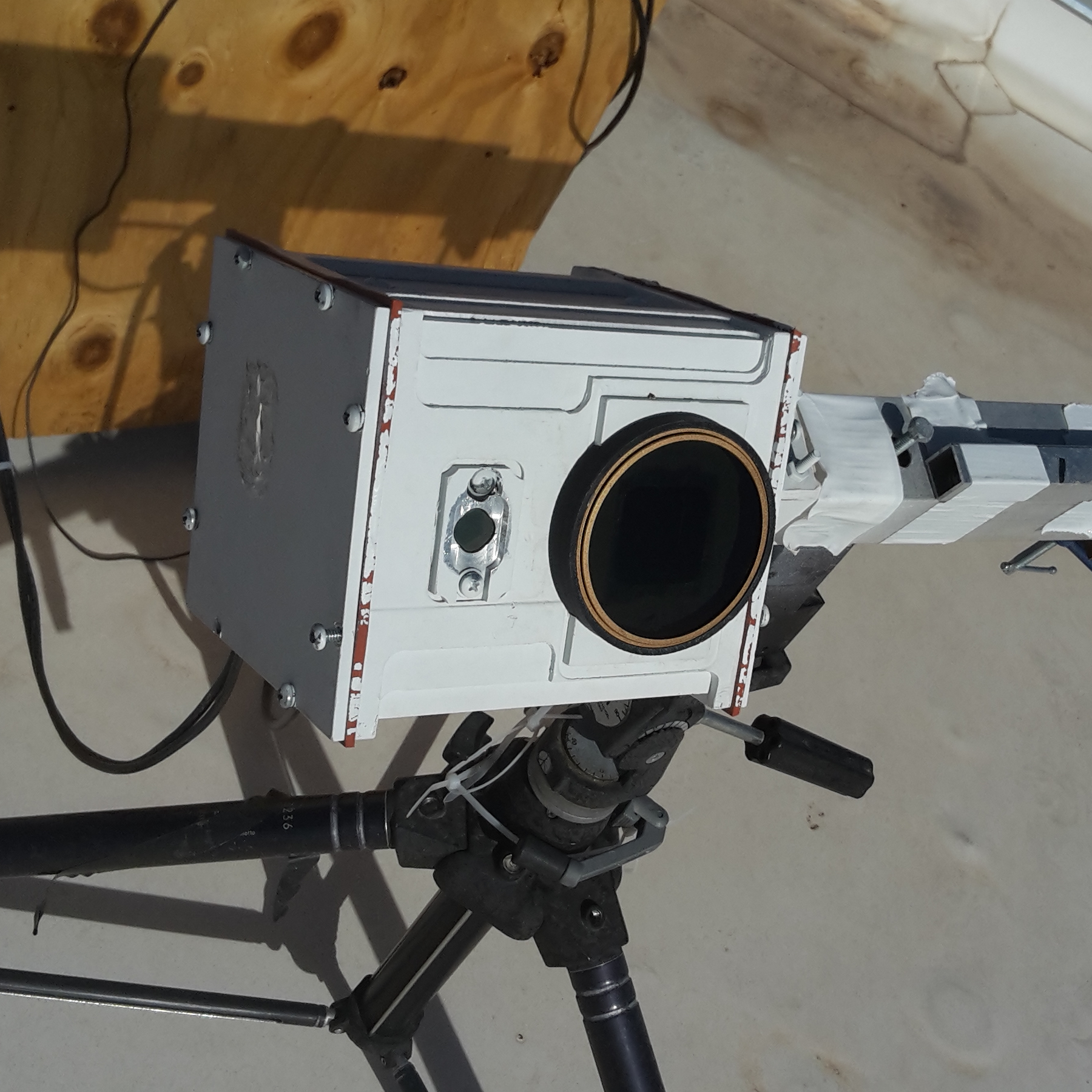}
    \end{subfigure}
    \begin{subfigure}{0.3275\textwidth}
        \centering
        \includegraphics[scale = 0.425]{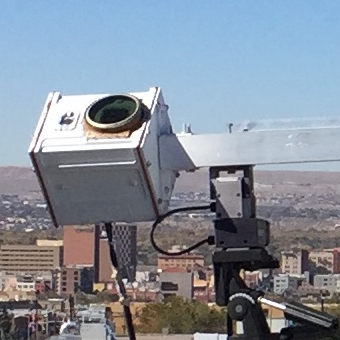}
    \end{subfigure}
    \caption{Germanium and neutral density lenses are attached to the IR and VI cameras, respectively, to filter the light beams and protect the cameras from weather hazards (left). The solar tracker's servomotors are mounted on the tripod and screwed to a metallic structure which counterbalances the enclosure's weight (middle). The sky-imaging system (right) installed on the roof of a UNM building.}
\label{fig:exterior}
\end{figure}

\subsubsection{Solar Tracker}

The device used for solar tracking is a commercial model manufacture by FLIR$\copyright$ and distributed by moviTHERM (Fig. \ref{fig:exterior}). The model is a PTU-E46-70. The tracking system is driven by two servomotors mounted to allow rotation on tilt and pan axes. The tracker is rated for a payload of 4.08 kg, (the maximum weight of the system). The rotational speed of the tracker is $60^\circ$ per second, and it has a resolution of $0.003^\circ$.
	
The degree of freedom of the tilt servomotor allows vertical maneuvers of $78^\circ$ with a vertical limit ranging from $-47^\circ$ to $+31^\circ$ with respect to the horizontal axis. The pan servomotor has more degrees of freedom. This allows the Sun to be tracked throughout the entire day. Its range is $\pm 159^\circ$. The motor speeds are adjustable up to $0.003^\circ$ per second. The tracker requires an input voltage ranging from 12 to 30 VDC. The corresponding  power is 13 W in maximum power mode, 6 W low-power mode and 1 W in holding power off mode. The system is controlled via Ethernet from a host computer. The control unit is connected to the servomotors via a DB-9 female connector.
	
Regarding the mechanical description of the tracker, the system weight is 1.36 kg, its dimensions are 7.6 cm height $\times$ 13 cm width $\times$ 10.17 cm depth. The control unit weights 0.227 kg, and its dimensions are 3.18 cm height $\times$ 8.26 cm width $\times$ 11.43 cm depth.
	
\subsubsection{Pyranometer}

As part of the DAQ, a pyranometer sensor has been designed and tested to provide the device with high portability and accuracy. Here,  the circuit design for signal conditioning of the pyranometer sensor (model LI-COR LI-200) is summarized. The prototype was designed to send data to a laptop, Raspberry Pi or any other computer through USB communication, to which a simple Python script needs to be added and executed. For portability purposes, the board we designed is provided with a 5 V power supply, so that a Raspberry Pi may also be connected to it. 

\begin{figure}[!htb]
    \begin{subfigure}{\textwidth}
        \centering
        \includegraphics[scale = 0.65]{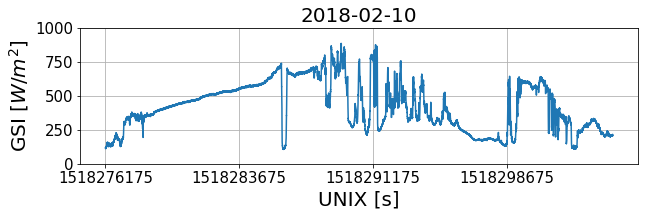}
    \end{subfigure}
    \begin{subfigure}{\textwidth}
        \centering
        \includegraphics[scale = 0.65]{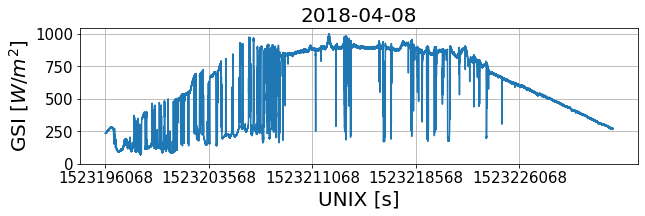}
    \end{subfigure}
    \caption{This figure shows two days of samples from the pyranometer. Day 2018-02-10 has 101510 samples and day 2018-04-08 has 131412 samples. The $x$ axis shows the UNIX time in seconds. The $y$ axis shows the GSI in $W/m^2$.}
\label{fig:pyranometer}
\end{figure}

The pyranometer design objective was to condition and measure the signal of the sensor by converting it to a 12-bit digital signal and saving the data by timestamp in a .csv file. The system is portable, easy to setup, and reliable. The pyranometer output voltage ranges from 0 to 20 mV, where 10 mV represents a radiation power density of 1,000 $W/m^2$ (see Fig. \ref{fig:pyranometer}). A 120 V power outlet is used to power the device. The system is connected to the Internet and has user log-in capabilities. 
	
Our design uses an Analog Devices AD629 instrumentation amplifier with 13 dB gain that conditions the signal to an ADC. An Atmel SAMD21G18 microcontroller is used to measure the signal output from the ADC and to transmit it by a USB port. The board is connected to a Raspberry Pi to save the signal's data in files. This Raspberry Pi is connected to the internet, so it is also able to transmit the data.

\subsection{Data Acquisition Software}

The tracking system and DAQ software is currently operative. The software was programmed in a single Python 2.7 script. The system is placed on the roof of the Electrical and Computer Engineering building at UNM central campus $(35.0821, -106.6259)$. The DAQ sessions can be visually monitored through a webpage. All devices were interconnected via a LAN network built by a DHCP server.

\subsubsection{Sun Position}

The Sun position is computed to update the servomotors’ position every second. The following set of equations was used to calculate azimuth and elevation angles\footnote{http://www.pveducation.org/}. These angles correspond to the servomotors’ pan and tilt axes. The tracking system must be aligned to the true North or South of the local geographic coordinates defined as $\left( \lambda, \phi \right)$, which are respectively longitude and latitude. The system must be leveled with respect to the horizontal and vertical axis. The first step is to calculate the Local Standard Time Meridian (LSTM), which is the time zone meridian with respect to the Greenwich meridian. The LSTM is computed as,
\begin{equation} 
	\mathrm{LSTM} = 15^\circ \cdot \Delta T_{GTM},
\end{equation}	
where $\Delta T_{GMT}$ is the difference between the Local Time (LT) and Greenwich Mean Time (GMT) and it is measured in hours. The equation of time (EoT) is an empirical equation that considers the eccentricity of earth's orbit and its axial tilt. It is expressed in minutes and is described by
\begin{equation} 
	\mathrm{EoT} = 9.87  \cdot  \sin(2\mathrm{B}) - 7.53 \cdot  \cos(\mathrm{G}) - 1.5  \cdot \sin(\mathrm{B}),
\end{equation}
where
\begin{equation} 
	\mathrm{B} = \frac{360}{365} \cdot \left( d - 81 \right).
\end{equation}
The units of $\mathrm{B}$ are degrees and $d$ corresponds to the number of days since the beginning of the year. Time Correction factor (TC) accounts for local effects that produce variations in the Local Solar Time(LST).These effects are caused by the eccentricity of the Earth’s orbit. TC is quantified in minutes, and its equation is
\begin{equation} 
	\mathrm{TC} = 4 \cdot \left( \lambda - \mathrm{LSTM} \right) + \mathrm{EoT},
\end{equation}
Constant 4 is related to the rotation of the earth, which rotates $1^{\circ}$ every 4 minutes, and $\lambda$ is the longitude. The LST is calculated by using the previous correction and Local Time (LT). The  following equation is expressed in hours
\begin{equation} 
	\mathrm{LST} = \mathrm{LT} + \frac{\mathrm{TC}}{60}.
\end{equation}
	
The Earth rotates $15^{\circ}$ per hour, so each hour away from the solar noon corresponds to an angular position. The Hour Angle (HRA) converts LST to the Sun's angle. Before noon, angles are negative and in the afternoon the angles obtained are positive. The expression for this is
\begin{equation} 
	\mathrm{HRA} = 15^\circ \cdot \left( \mathrm{LST} - 12 \right).
\end{equation}
	
The Sun's declination angle $\delta$ varies seasonally, and it accounts for the angular difference between the Equator and the Earth's center. $\delta$ is obtained from formula
\begin{equation} 
	\delta = 23.45^\circ \cdot \sin \left[ \frac{360}{365} \cdot \left( d - 8 \right) \right],
\end{equation}
where $d$ represents the day since the beginning of the year. The rotations on the tracker’s pan and tilt axes are provided by the Elevation $\epsilon$ and Azimuth $\alpha$ angles. The Elevation angle quantifies the Sun's angular height, and the azimuth $\alpha$ is the Sun's angular position on the horizon. The angles are calculated from the following equations,
\begin{equation} 
    \begin{split}
        \epsilon &= \sin^{-1} \left[\sin\delta  \cdot \sin\phi + \delta \cdot \cos\phi \cdot \cos(\mathrm{HRA}) \right], \\
	    \alpha &= \cos^{-1} \left[ \frac{\sin\delta \cdot \cos\varphi - \cos\delta \cdot \sin\varphi \cdot \cos(\mathrm{HRA})}{\cos \xi } \right],
    \end{split}
\end{equation}
where $\mathrm{HRA}$ is the hour angle, $\phi$ is latitude, and 
\begin{equation} 
    \xi = \sin^{-1} \left[ \sin\delta  \cdot \sin\phi + \delta \cdot \cos\phi \cdot \cos(\mathrm{HRA}) \right].
\end{equation}
Alternatively, $\xi$ can be also calculated as $\xi = 90^\circ + \phi - \delta$. The zenith angle $\zeta$ is the difference between the elevation angle $\epsilon$ and the vertical axis, so that
\begin{equation} 
	\zeta = 90^\circ - \epsilon.
\end{equation}
	
The tracking system initializes one hour after the sunrise and stops one hour before sunset, so that it is inactive during the hours when the Sun’s elevation is too low to generate energy by PV systems. Inverters require a minimum amount of power generation, which is reached when Sun's position is above and below these time limits. The sunrise (SR) and sunset (SS) are expressed in decimal hours and they are calculated in the following equations
\begin{equation}
\begin{split}
	\mathrm{SR} &= 12 - \frac{1}{15^\circ}  \cdot\cos^{-1} \left(\frac{-\sin\phi \cdot \sin\delta}{\cos\phi \cdot \cos\delta}\right) - \frac{ \mathrm{TC} }{60}, \\
	\mathrm{SS} &= 12 + \frac{1}{15^\circ} \cdot \cos^{-1} \left(\frac{-\sin\phi \cdot \sin\delta}{\cos\phi  \cdot \cos\delta} \right) - \frac{ \mathrm{TC} }{60}.
\end{split}
\end{equation}

\subsubsection{Noise Attenuation}\label{sec:322}
\label{averaged_image}

\begin{figure}[!htb]
    \centering
    \begin{subfigure}{0.45\textwidth}
        \includegraphics[scale = 0.25]{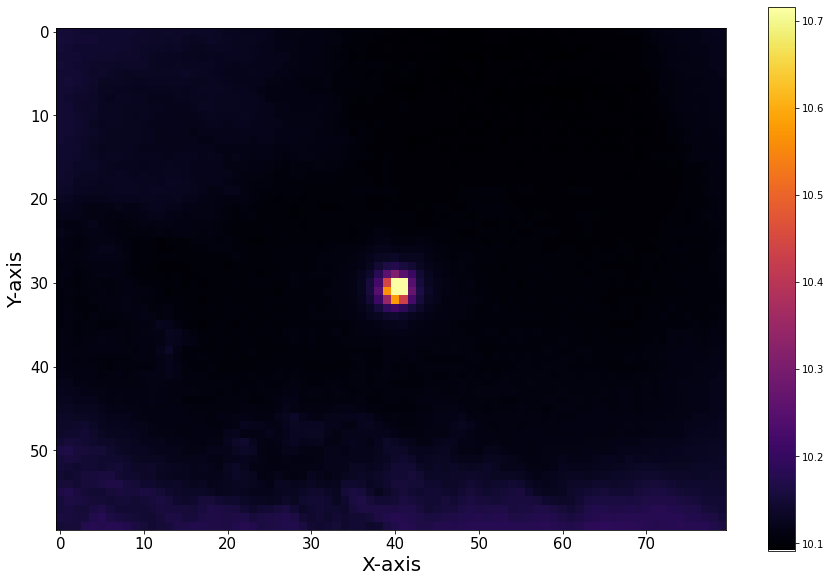}
    \end{subfigure}
    \begin{subfigure}{0.45\textwidth}
        \includegraphics[scale = 0.25]{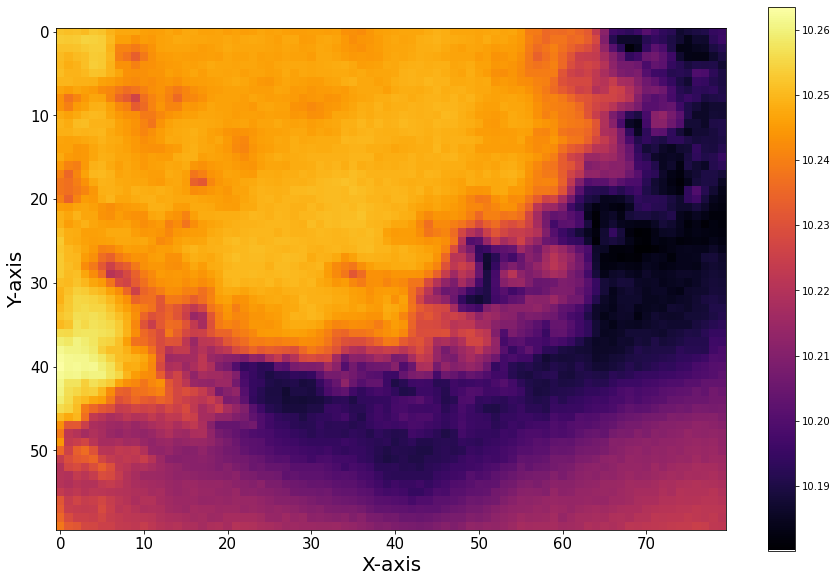}
    \end{subfigure}
    \begin{subfigure}{0.45\textwidth}
        \includegraphics[scale = 0.25]{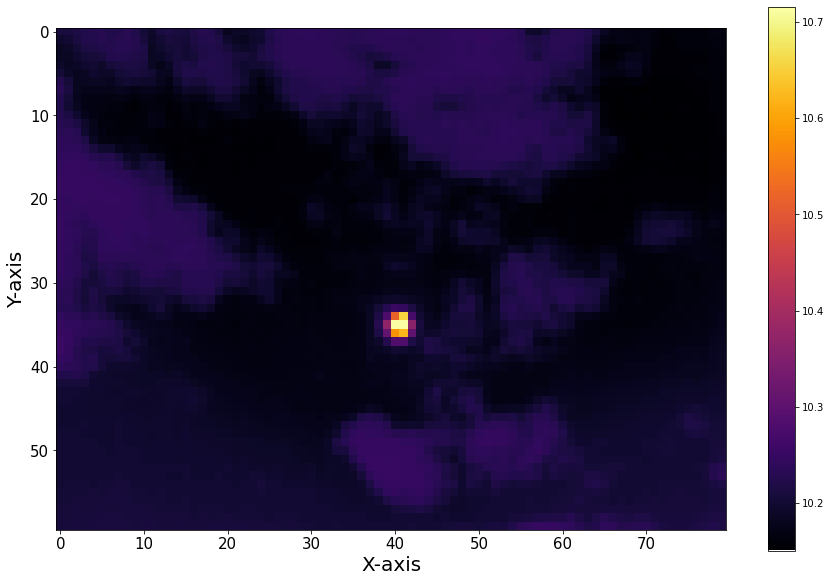}
    \end{subfigure}
    \begin{subfigure}{0.45\textwidth}
        \includegraphics[scale = 0.25]{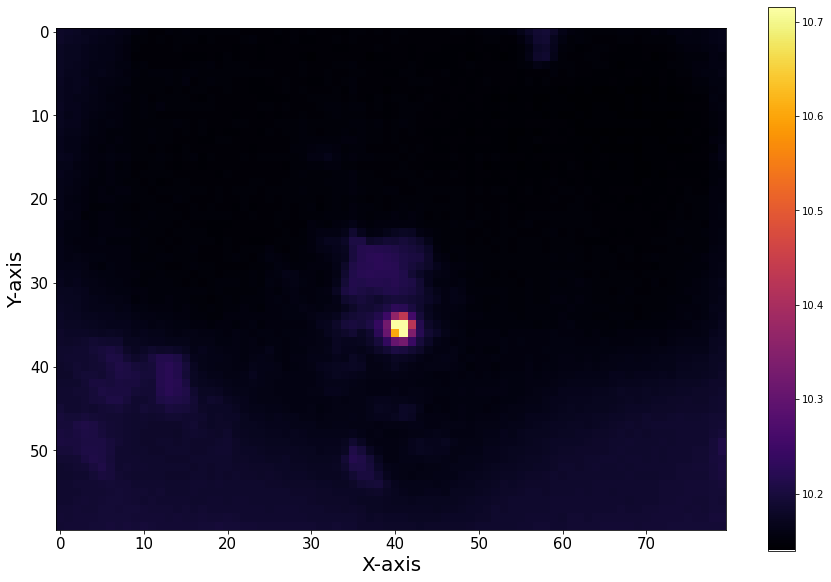}
    \end{subfigure}
    \caption{This figure shows samples (stored in the repository) of IR images that were acquired in different days. The intensity of the pixels in the images is displayed in logarithmic scale. The resolution of the images is $60 \times 80$ and their sample rate is four images per minute.}
    \label{fig:infrared}
\end{figure}

The captures from the IR cameras are noisy. Therefore, we propose to attenuate the noise by averaging $N=10$ frames taken at capture $k$, with a frame rate of 9 images per second. The camera frame rates are lower than the computer socket reading speed, so we need to assess whether each acquired frame read from the buffer is new or if it has been previously acquired in our set. Also, it is possible that a capture was defective. To detect these situations, we first acquire $N$ consecutive images and we compute the sample mean image of the set $\bar{\mathbf{I}} = \frac{1}{N} \cdot \sum_{i = 1}^N \mathbf{I}_i$. Then, for all images in the set, we compute the Pearson coefficient 
\begin{equation}
    \rho_{i,j}  = \frac{ ( {\bf I}_i - \bar{\bf I} ) \cdot ( {\bf I}_j - \bar{\bf x} )}{ \sqrt{(\mathbf{\mathbf{x}}_i - \bar{\mathbf{I}} )^2} \cdot \sqrt{ ( {\bf I}_j - \bar{\bf I} )^2}} , \quad \forall i,j = 1, \dots, N.
\end{equation}
If a pair of frames has a coefficient equal to one, we discard one of them since the frames are the same. If an image has a coefficient less than $0.9$ with respect the rest, we assume that the frame is defective, and it is also discarded. The mean is recomputed with the remaining frames and then stored (see Fig. \ref{fig:infrared}). 

\subsubsection{Exposure Switching}

The scattering effect produced by solar irradiance in the VI images precludes  the detection of the clouds in the circumsolar region (when the exposure time is long). The light in the outer region of the image is faint when the exposure time is short. To resolve this issue, we propose to average the captured images at different exposure times. 
A frame is taken with four different exposure times equal to $1$ ms, $4$ ms, $12$ ms and $28$ ms. The process is repeated 10 times, and the frames with same exposure time are averaged using the procedure in Subsection \ref{sec:322}, which results in 4 different frames corresponding to the four different exposures. This images are further fused using the procedure below.  

\subsubsection{Visible Image Fusion}

We introduce a method to fuse images with different exposure times, generating a High Dynamic Range (HDR) of 16 bits, and combining cloud information from the  different images.  The frame is defined as $I^k_{e,c}$, $1\leq e \leq 4$ as each one of the RGB components $1 \leq c \leq 3$ of image $I_e^k$ at instant $k$ and exposure $T_e \in \{1,4, 12, 28\}$ ms. To merge the images we first regularized the RGB components in each exposition to avoid division by zero when light is too dim,
\begin{equation}
    \mathbf{I}^k_{e, c} + \lambda, \quad \mathbf{I}^k_{e, c} \in \mathbb{R}^{D \times D}.
\end{equation}
We convert the images to grayscale by a weighted sum of the image RGB channels with the corresponding \emph{Luma} coding system coefficients  $\beta_c = \{0.299, 0.587, 0.114 \}$ \cite{Poynton2012}, 
\begin{equation}
    \mathbf{I}^k_e = \sum_{c = 1}^C \mathbf{I}^k_{e, c} \cdot \beta_c
\end{equation}

The proposed method is based on the distance from a pixel to the Sun. We know that the Sun is always centered in the frames. Therefore, we define our fusion mask as a binary valued frame function of the radial distance of the Sun to a pixel, such as
\begin{equation}
    \mathcal{M} \left( r\right) \triangleq \left\{{\bf M}_{i,j}=\mathbb{I} \left( \left( x_i - x_{sun} \right)^2 + \left( y_j - y_{sun} \right)^2 \leq r^2 \right)  \right\},
\end{equation}
where $x_i$ and $y_j$ are coordinates of a frame and $\{x_{sun}, y_{sun} \}$ are the coordinates of the Sun. We define a mask $\mathbf{M}_e$ for each exposition time corresponding to  a radius,
\begin{equation}
     \mathbf{M}_e  = \mathcal{M}  \left( r_e \right), \quad \forall r_e = \{5, 6.25, 12.5, 25\}
\end{equation}
the set of fusion masks is then $\mathbf{M}_e = \{\mathbf{0}, \mathbf{M}_2, \cdots, \mathbf{M}_E, \mathbf{1}\}$. We convolve each one of the masks with a Gaussian kernel,
\begin{equation}
    \mathcal{G} \left( x, y\right) = \frac{1}{C}\exp \left\{-\frac{x^2 - y^2}{2\sigma^2} \right\},
\end{equation}
where $\sigma=7.5$, $C$ is a normalization constant, so the kernel integrates to 1,  and has dimensions $15 \times 15$. This operation is done to blur the mask pixels in the edges with the objective of smoothing out the transition between masked image regions in the resulting merged image. We define $\tilde{\bf M}_e$ as each one of the masks that were processed using the Gaussian filter. 

The different frames are then merged with the  formula   
\begin{equation}
   \mathbf{X}^k = \alpha_1^k \cdot \left( \tilde{\bf M}_1 \odot {\bf I}_1^k \right) + \sum_{e = 2}^4 \frac{\alpha^k_e}{e} \cdot \left[ \tilde{\mathbf{M}}_{e-1} \odot \left( \mathbf{1} - \tilde{\mathbf{M}}_{e}\right) \odot \cdot \mathbf{I}^k_e \right], \quad \mathbf{X}^k \in \mathbb{R}^{D \times D}
\end{equation}
where the Hadamard or element-wise product $\odot$ is used to create the concentric rings with radii $r_{e - 1}$ and $r_e$. Coefficients $\alpha_e^k$ are used to force the intensity of pixels of the outer edge of each ring to be equal to the intensities of the inner edge of the next ring. We obtain a set of weights $\alpha^k_e = \{1, \alpha^k_2, \cdots, \alpha^k_E\}$ for each frame $k$ by averaging the pixels of each frame outer edge as 
\begin{equation}
    \alpha^k_{e + 1} = \alpha^k_e \cdot \frac{\sum_i \sum_j \mathbf{I}^k_e \odot \mathbf{R}^2_e }{\sum_i \sum_j \mathbf{I}^k_{e + 1} \odot \mathbf{R}^1_e }, \quad \forall e \in \{ 1, \cdots, E - 1 \}, \quad \alpha^k_{e}  \in \mathbb{R}.
\end{equation}
where $\mathbf{R^1_e}$ and $\mathbf{R^2_e}$ are rings of radius $\epsilon = \sqrt{2}$ defined as   
\begin{equation}
    \begin{split}  
        \mathbf{R}^1_e &= \mathcal{M}\left( r_e \right) \oplus \mathcal{M}  \left( r_e + \epsilon \right) , \quad \forall r_e = \{r_1, \cdots, r_E\}\\
        \mathbf{R}^2_e &= \mathcal{M}\left( r_e \right) \oplus \mathcal{M}  \left( r_e - \epsilon \right) , \quad \forall r_e = \{ r_1, \cdots, r_E\}.
    \end{split}
\end{equation}

\begin{figure}[!htb]
    \begin{subfigure}{0.245\textwidth}
        \centering
        \includegraphics[scale = 0.185]{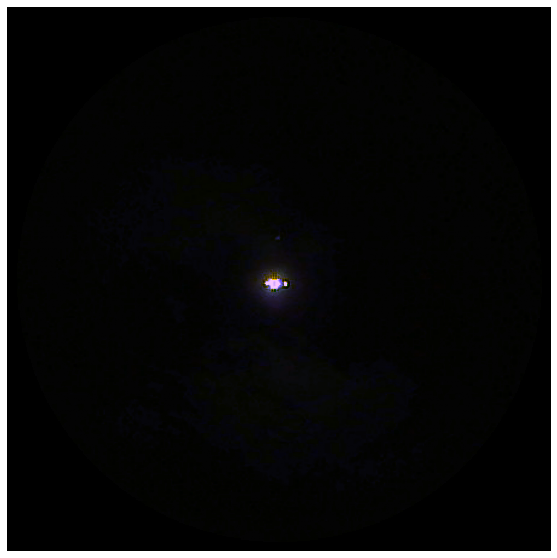}
        \includegraphics[scale = 0.185]{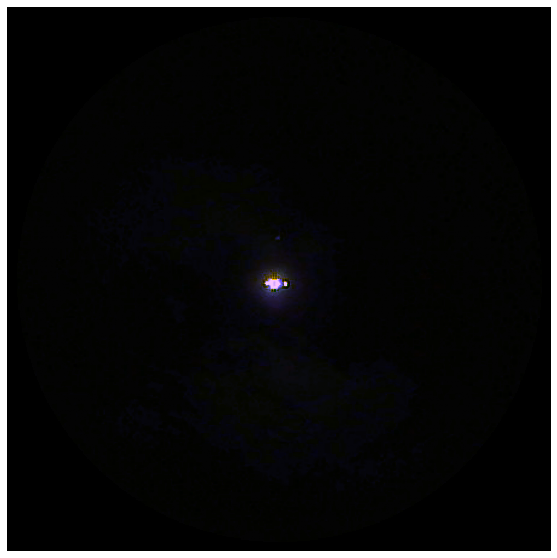}
        \includegraphics[scale = 0.185]{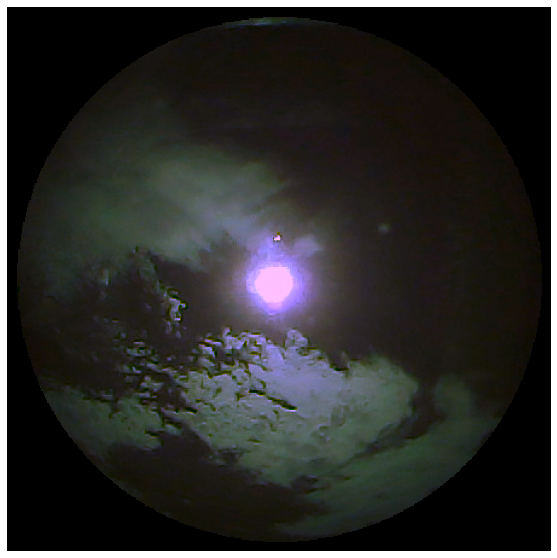}
        \includegraphics[scale = 0.185]{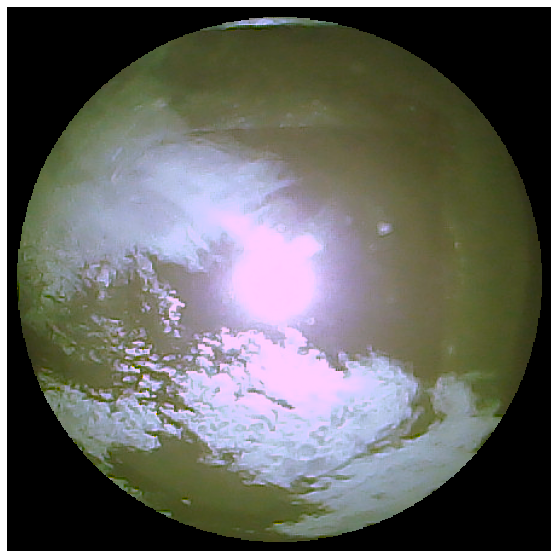}
        \includegraphics[scale = 0.185]{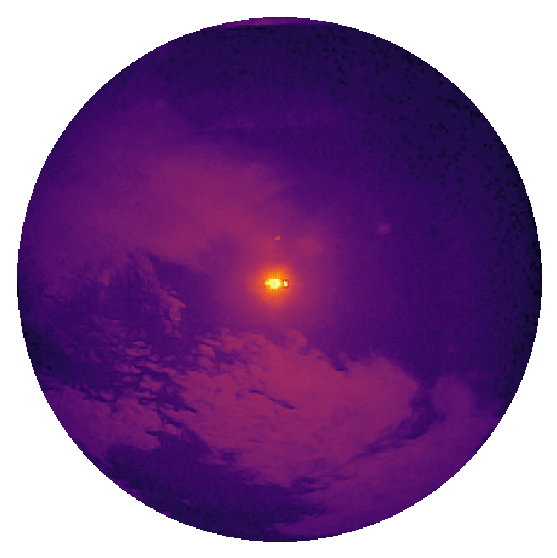}
    \end{subfigure}
    \begin{subfigure}{0.245\textwidth}
        \centering
        \includegraphics[scale = 0.185]{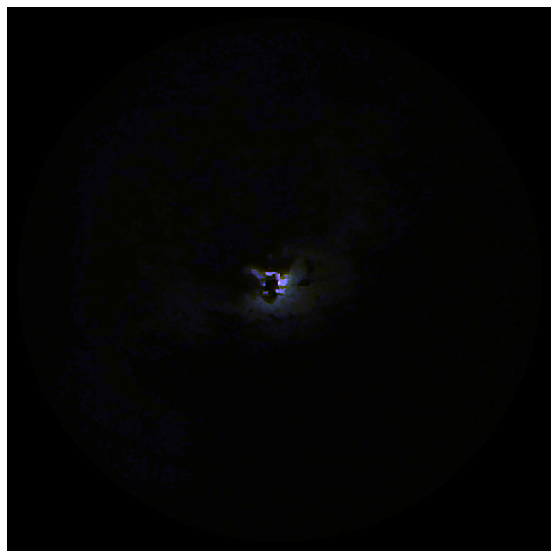}
        \includegraphics[scale = 0.185]{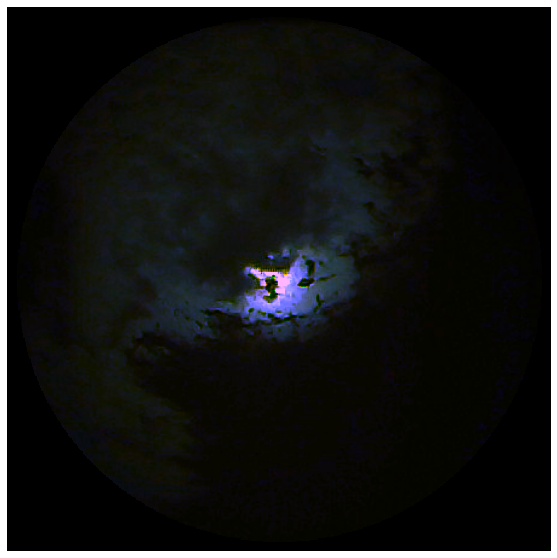}
        \includegraphics[scale = 0.185]{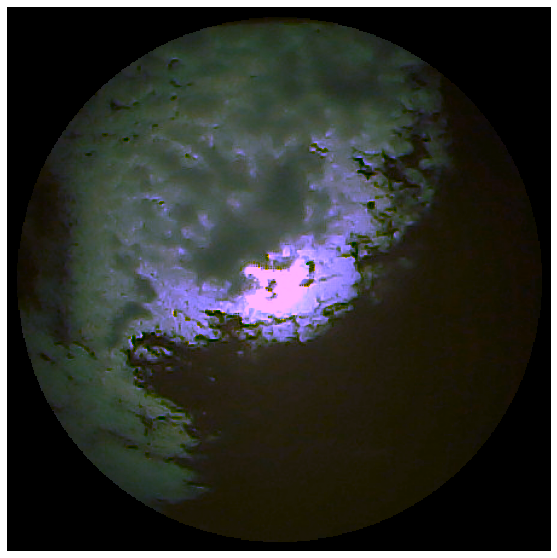}
        \includegraphics[scale = 0.185]{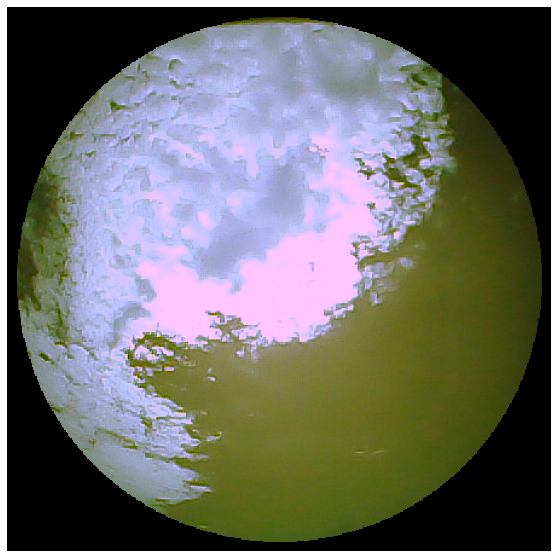}
        \includegraphics[scale = 0.185]{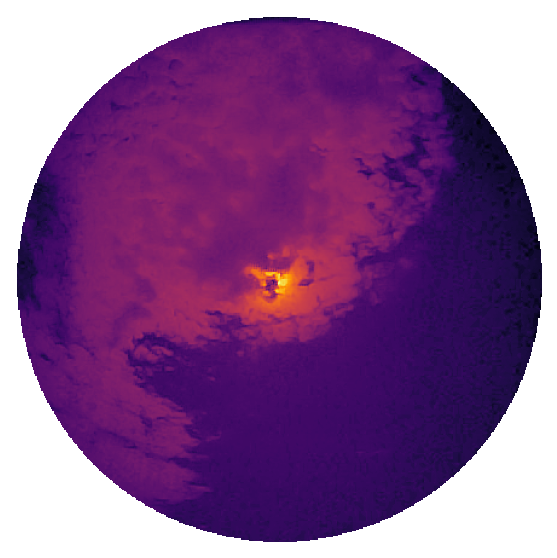}
    \end{subfigure}
    \begin{subfigure}{0.245\textwidth}
        \centering
        \includegraphics[scale = 0.185]{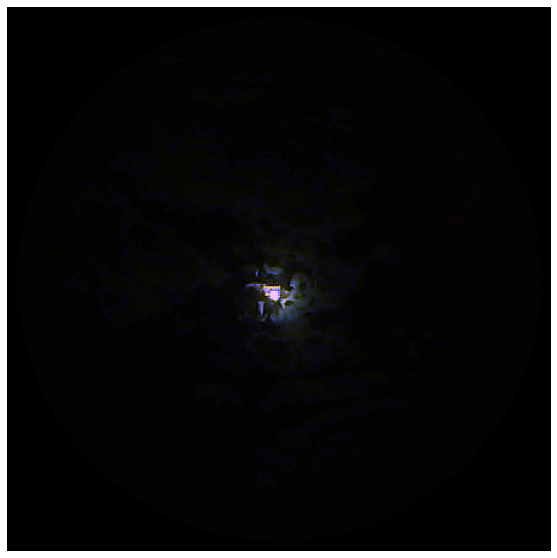}
        \includegraphics[scale = 0.185]{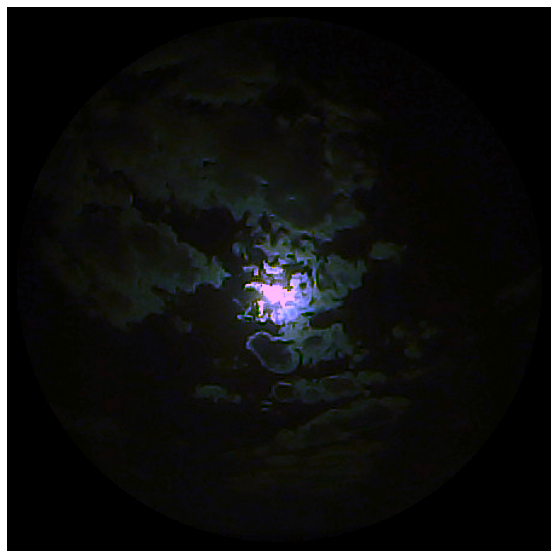}
        \includegraphics[scale = 0.185]{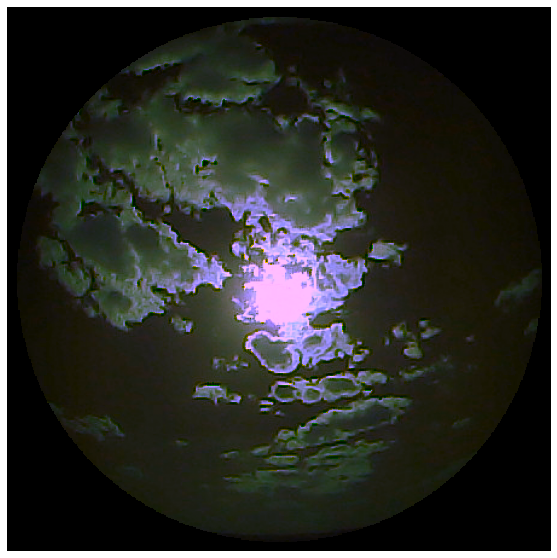}
        \includegraphics[scale = 0.185]{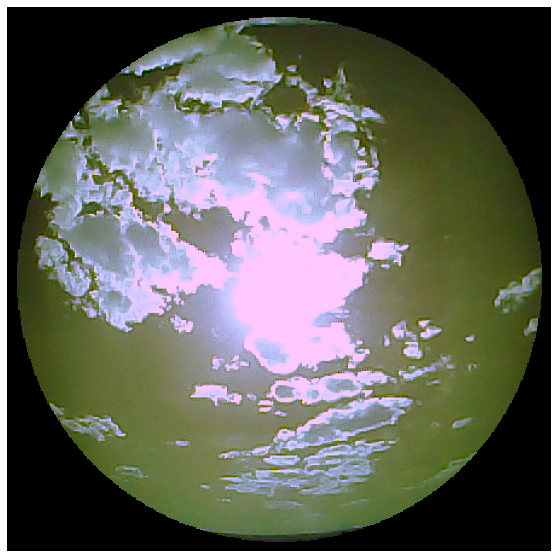}
        \includegraphics[scale = 0.185]{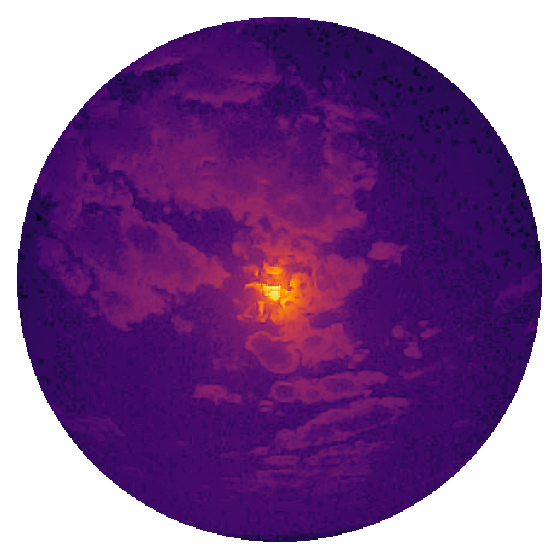}
    \end{subfigure}
    \begin{subfigure}{0.245\textwidth} 
        \centering
        \includegraphics[scale = 0.185]{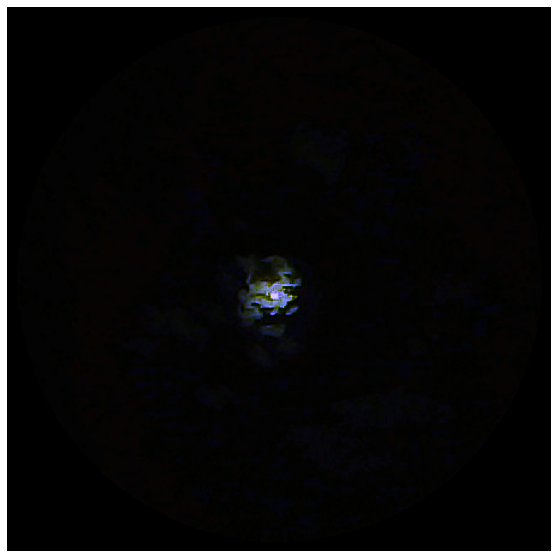}
        \includegraphics[scale = 0.185]{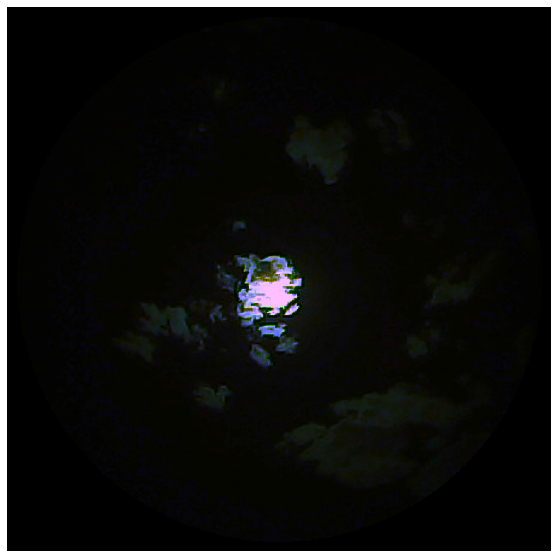}
        \includegraphics[scale = 0.185]{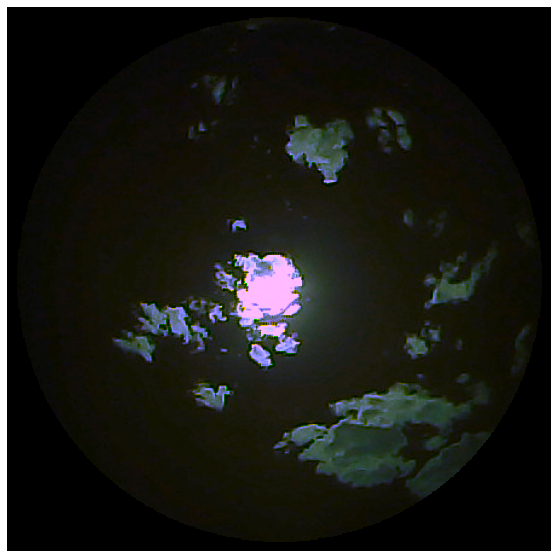}
        \includegraphics[scale = 0.185]{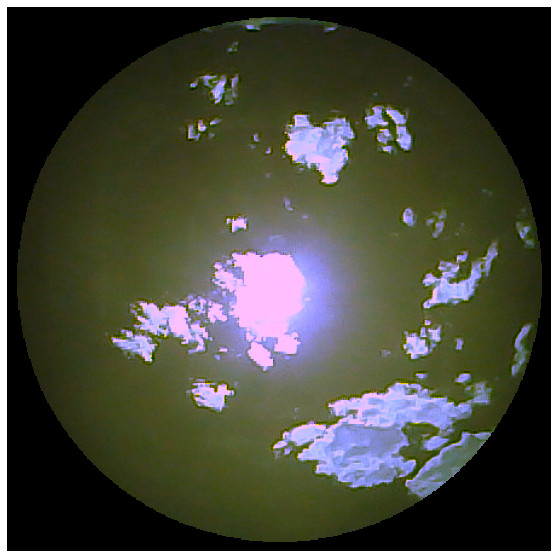}
        \includegraphics[scale = 0.185]{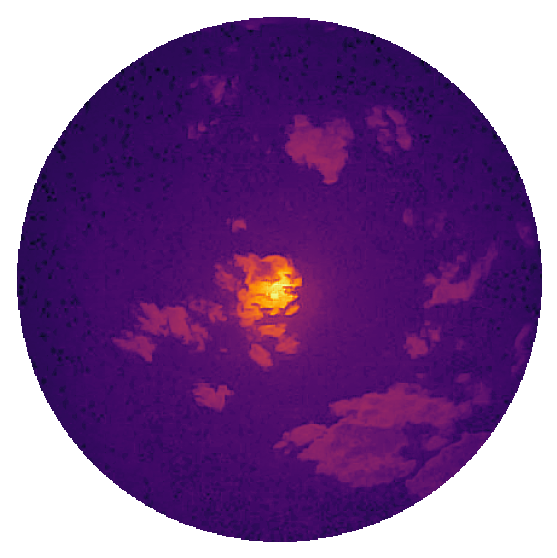}
    \end{subfigure}
    \caption{Result of the image fusion procedure. The four first rows of each column show the average of 10 sky images captured at a given exposure time (from top to bottom, 1, 4, 12 and 28 ms). The last row shows the images after completing the fusion algorithm. The resulting images are displayed in logarithmic-scale because of the high intensity of the pixels in the circumsolar region. Their size is $450 \times 450$. Pixels without information are not displayed.}
    \label{fig:exposure}
\end{figure}

The images are then converted to 16 bit arithmetic for further processing, and stored. Since the maximum pixel amplitude after the fusion is 225, the 8 to 16 bit conversion equation is,
\begin{align}
    \mathbf{I}^k = \left( \frac{\mathbf{X}^k}{225} \right) \cdot 2^{16}, \quad \mathbf{I}^k \in \mathbb{R}^{D \times D}.
\end{align}

The outer circular region in the fused image is set to an intensity value of 0, as it displays artifacts produced by its own lens holder and case (see Fig. \ref{fig:exposure}). Therefore, it does not include relevant information for the prediction.

\section*{Acknowledgments}

This work has been supported by NSF EPSCoR grant number OIA-1757207 and the King Felipe VI endowed Chair. Authors would like to thank the UNM Center for Advanced Research Computing, supported in part by the National Science Foundation, for providing the high performance computing and large-scale storage resources used in this work.

\bibliographystyle{unsrt}  
\bibliography{mybibfile}

\end{document}

%% file: REVISED_table.tex
{\fontsize{7.5pt}{9pt}\selectfont
\begin{longtable}{|p{33mm}|p{94mm}|}
\hline
\endhead
\hline
\endfoot
Subject                & Energy Engineering and Power \\
\hline                         
Specific subject area  & Artificial intelligence system to forecast solar energy in Micro Grids and Smart Grids powered by photovoltaic technology \cite{ZAPPA2019}. \\
\hline
Type of data           & 
Image (.png) \newline
Table (.csv)\\             
\hline
How data were acquired & 
                         Instruments: Infrared Camera, Visible Camera, Solar Tracker, Pyranometer sensor, Raspberry Pi, Motherboard \newline
                         Make and model and of the instruments used: FLIR$\copyright$ Lepton 2.5 infrared camera mounted on Pure Thermal 1 board manufactured by Group Gets. ELP$\copyright$ visible camera model 8541707515. Tracker FLIR$\copyright$ model PTU-E46-70. LI-COR$\copyright$  Pyranometer sensor model LI-200. Raspberry Pi 2B manufactured by SONY$\copyright$. ASRock$\copyright$ model Syper Alloy J3455-ITX Quad-core. \\

\hline                         
Data format            & Raw \\
\hline                         
Parameters for         
data\newline 
collection             & The sampling interval of the cameras is 15 seconds when the Sun's elevation angles is $>15^\circ$. There are approximately 1,200 to 2,400 captures per day from each camera depending of the year day. The GSI signal has a sampling rate ranging from 4 to 6 samples per second. The Sun position files are generate out of the time recorded in the pyranometer files. The weather station data is updated every 10 minutes. \\  

\hline
Description of          
data\newline 
collection             & The tracking system and DAQ software is currently operative. The software was programmed in Python 2.7. The system is placed on the top roof area of the Mechanical Engineering building in UNM central campus. The DAQ session can be visually monitored through a webpage. All devices are interconnected via a LAN network built by DHCP server. The weather station is located at the University of New Mexico Hospital, and both its real-time and historical data are publicly accessible\footnote{https://www.wunderground.com/dashboard/pws/KNMALBUQ473}. \\
\hline                         
Data source location   & 
                         Institution: The University of New Mexico \newline
                         City/Town/Region: Albuquerque, New Mexico \newline
                         Country: United States of America \newline
                         Latitude and longitude for collected samples: 35.0821, -106.6259 \\
\hline                         
\hypertarget{target1}
{Data accessibility}   & 
                        Repository name: Girasol, a Sky Imaging and Global Solar Irradiance Dataset \newline
                        Data identification number: 10.5061/dryad.zcrjdfn9m\newline
                        Direct URL to data and instructions for accessing these data: \url{https://doi.org/10.5061/dryad.zcrjdfn9m} \\                         
\hline                         
Related                 
research\newline
article                & G. Terr\'en-Serrano, M. Mart\'inez-Ram\'on, Multi-Layer Wind Velocity Field Visualization in Infrared Images of Clouds for Solar Irradiance Forecasting, Applied Energy. In Press. \\
      
\end{longtable}}

%% file: arxiv.bbl
\begin{thebibliography}{10}

\bibitem{YANG2018}
Dazhi Yang, Jan Kleissl, Christian~A. Gueymard, Hugo~T.C. Pedro, and
  Carlos~F.M. Coimbra.
\newblock History and trends in solar irradiance and pv power forecasting: A
  preliminary assessment and review using text mining.
\newblock {\em Solar Energy}, 168:60 -- 101, 2018.
\newblock Advances in Solar Resource Assessment and Forecasting.

\bibitem{MAMMOLI2019}
Andrea Mammoli, Guillermo Terren-Serrano, Anthony Menicucci, Thomas~P Caudell,
  and Manel Mart{\'\i}nez-Ram{\'o}n.
\newblock An experimental method to merge far-field images from multiple
  longwave infrared sensors for short-term solar forecasting.
\newblock {\em Solar Energy}, 187:254--260, 2019.

\bibitem{DIAGNE2013}
Maimouna Diagne, Mathieu David, Philippe Lauret, John Boland, and Nicolas
  Schmutz.
\newblock Review of solar irradiance forecasting methods and a proposition for
  small-scale insular grids.
\newblock {\em Renewable and Sustainable Energy Reviews}, 27(Supplement C):65
  -- 76, 2013.

\bibitem{CHENG2017b}
Hsu-Yung Cheng.
\newblock Cloud tracking using clusters of feature points for accurate solar
  irradiance nowcasting.
\newblock {\em Renewable Energy}, 104:281 -- 289, 2017.

\bibitem{FURLAN2012}
Claudia Furlan, Amauri~Pereira {de Oliveira}, Jacyra Soares, Georgia Codato,
  and João~Francisco Escobedo.
\newblock The role of clouds in improving the regression model for hourly
  values of diffuse solar radiation.
\newblock {\em Applied Energy}, 92:240 -- 254, 2012.

\bibitem{KONG2020}
Weicong Kong, Youwei Jia, Zhao~Yang Dong, Ke~Meng, and Songjian Chai.
\newblock Hybrid approaches based on deep whole-sky-image learning to
  photovoltaic generation forecasting.
\newblock {\em Applied Energy}, 280:115875, 2020.

\bibitem{CHOU2019}
Jui-Sheng Chou and Ngoc-Son Truong.
\newblock Cloud forecasting system for monitoring and alerting of energy use by
  home appliances.
\newblock {\em Applied Energy}, 249:166 -- 177, 2019.

\bibitem{Deng2019}
C.~{Deng}, Z.~{Li}, W.~{Wang}, S.~{Wang}, L.~{Tang}, and A.~C. {Bovik}.
\newblock Cloud detection in satellite images based on natural scene statistics
  and gabor features.
\newblock {\em IEEE Geoscience and Remote Sensing Letters}, 16(4):608--612,
  April 2019.

\bibitem{Ye2019}
L.~{Ye}, Z.~{Cao}, Y.~{Xiao}, and Z.~{Yang}.
\newblock Supervised fine-grained cloud detection and recognition in whole-sky
  images.
\newblock {\em IEEE Transactions on Geoscience and Remote Sensing},
  57(10):7972--7985, Oct 2019.

\bibitem{VOYANT2017}
Cyril Voyant, Gilles Notton, Soteris Kalogirou, Marie-Laure Nivet, Christophe
  Paoli, Fabrice Motte, and Alexis Fouilloy.
\newblock Machine learning methods for solar radiation forecasting: A review.
\newblock {\em Renewable Energy}, 105(Supplement C):569 -- 582, 2017.

\bibitem{GARCIA2018}
O.~García-Hinde, G.~Terrén-Serrano, M.Á. Hombrados-Herrera,
  V.~Gómez-Verdejo, S.~Jiménez-Fernández, C.~Casanova-Mateo, J.~Sanz-Justo,
  M.~Martínez-Ramón, and S.~Salcedo-Sanz.
\newblock Evaluation of dimensionality reduction methods applied to numerical
  weather models for solar radiation forecasting.
\newblock {\em Engineering Applications of Artificial Intelligence}, 69:157 --
  167, 2018.

\bibitem{CRESPI2018}
Francesco Crespi, Andrea Toscani, Paolo Zani, David Sánchez, and Giampaolo
  Manzolini.
\newblock Effect of passing clouds on the dynamic performance of a csp tower
  receiver with molten salt heat storage.
\newblock {\em Applied Energy}, 229:224 -- 235, 2018.

\bibitem{CHEN2020}
Xiaoyang Chen, Yang Du, Enggee Lim, Huiqing Wen, Ke~Yan, and James Kirtley.
\newblock Power ramp-rates of utility-scale pv systems under passing clouds:
  Module-level emulation with cloud shadow modeling.
\newblock {\em Applied Energy}, 268:114980, 2020.

\bibitem{ZAPPA2019}
William Zappa, Martin Junginger, and Machteld {van den Broek}.
\newblock Is a 100\% renewable european power system feasible by 2050?
\newblock {\em Applied Energy}, 233-234:1027 -- 1050, 2019.

\bibitem{Poynton2012}
Charles Poynton.
\newblock {\em Digital Video and HD: Algorithms and Interfaces}.
\newblock Morgan Kaufmann Publishers Inc., San Francisco, CA, USA, 2 edition,
  2012.

\end{thebibliography}
